\documentstyle[12pt]{article}

\def\unlock{\catcode`@=11 }
\def\lock{\catcode`@=12 }
\unlock
\def\@@eqncr{\let\@tempa\relax\global\advance\@eqcnt by \@ne
    \ifcase\@eqcnt \def\@tempa{& & & &}\or \def\@tempa{& & &}\or
     \def\@tempa{& &}\or \def\@tempa{&}\else\fi
     \@tempa \if@eqnsw\@eqnnum\stepcounter{equation}\fi
     \if@defeqnsw\global\@eqnswtrue\else\global\@eqnswfalse\fi
     \global\@eqcnt\z@\cr}

\def\@eqnacr{{\ifnum0=`}\fi\@ifstar{\@yeqnacr}{\@yeqnacr}}
\def\@yeqnacr{\@ifnextchar [{\@xeqnacr}{\@xeqnacr[\z@]}}
\def\@xeqnacr[#1]{\ifnum0=`{\fi}\cr \noalign{\vskip\jot\vskip #1\relax}}
\def\eqalign{\null\,\vcenter\bgroup\openup1\jot \m@th \let\\=\@eqnacr
\ialign\bgroup\strut
\hfil$\displaystyle{##}$&$\displaystyle{{}##}$\hfil\crcr}
\def\endeqalign{\crcr\egroup\egroup\,}

\lock

\catcode`\@=11
\font\tenmsb=msbm10
\font\sevenmsb=msbm7 \font\fivemsb=msbm5 \newfam\msafam
\newfam\msbfam
\textfont\msbfam=\tenmsb
\scriptfont\msbfam=\sevenmsb \scriptscriptfont\msbfam=\fivemsb

\def\Bbb{\ifmmode\let\next\Bbb@\else
\def\next{\errmessage{Use \string\Bbb\space only in math
mode}}\fi\next}
\def\Bbb@#1{{\Bbb@@{#1}}} \def\Bbb@@#1{\fam\msbfam#1}
\catcode`\@=12

\newcommand{\x}{\mathop{\mbox{$\lefteqn{\times} \dot{\hphantom{-}}$}}}
\catcode `\@=11
\@addtoreset{equation}{section}

\catcode `\@=12
 \date{}
\title{Group Theoretical Examination of the Relativistic Wave Equations
on Curved Spaces.  \\ I.Basic Principles}
\author{S.A.Pol'shin\thanks{E-mail: semyon.a.polshin@univer.kharkov.ua} \\
\normalsize\it  Department of Physics, Kharkov State University, \\
\normalsize\it  Svobody Sq., 4, 310077, Kharkov, Ukraine }
\begin{document}
\maketitle
\thispagestyle{empty}
\begin{abstract}
The basic principles of generalization of the group theoretical approach to
the relativistic wave equations on curved spaces are examined. The general
method of  the determination of wave equations from the known symmetry group
of a symmetrical curved space is described.  The method of obtaining the
symmetrical spaces in which  invariant wave equations admit the limiting
passage  to the relativistic wave equations on the flat (not necessarily
real) spaces is explained.  Starting from the equations for  massless
particles and from the Dirac equation in  the Minkowski space, admissible real

symmetrical spaces are founded. The mentioned procedure is carried out also
for the complex spaces. As the basic point, the wave equations on the flat
complex space are considered. It is shown that the usual Dirac equation
written down in the complex variables does not lead to any curved space. By
"decomposition" of the mentioned form of the Dirac equation the equation
leading to the space ${\Bbb C}{\Bbb P}^{n}$ and being alternative to the
usual Dirac equation on ${\Bbb C}^{n}$ is constructed.
\end{abstract}

\section {Introduction} \label {I.1}
For the long time  the connection of elementary particles with various
classes of irreducible representations of the Minkowski space  symmetry
group,  i.e the Poincar\'e group~\cite{18} is well known. Also, it is well
 known, that it  is possible to deduce  from the fixed representation of the
 Poincar\'e group the wave equations, which represent restrictions imposed on
wave functions for they to belong to the space of the appropriate irreducible
representation of the Poincar\'e group~\cite {19,1,3}.
  Thus, one can deduce the wave equations for elementary
particles from  properties of the space-time symmetry group only, without
 additional suppositions.

However, actually, this approach has a much more generality and is applicable
to wave equations in symmetric curved spaces. Earlier
attempts to its generalization on curved spaces in one special case were
already studied $-$ in the case of de Sitter and Anti-de Sitter spaces; it
is possible to note~\cite{23}  spin zero particles,~\cite{7,32,36}
massive particles of spin 1/2 and~\cite{20,46} particles of higher spins. The
physical importance of the de Sitter space  is explained by that it, unique
of all real four-dimensional curved spaces (excepting the Anti-de Sitter
space), has the full 10-parametric symmetry group and, consequently, is the
favorite polygon in the quantum field theory in curved space-time
(see~\cite{24,33} and references therein).  Invariant wave equations in the
${\Bbb R}^{1}\otimes S^3$ space have also been treated in the
M.Carmeli's series of papers (see~\cite{42} and references trerein).

In the present paper the basic principles of  the group theoretical approach
to the relativistic wave equations generalization on curved spaces (generally
speaking, they are not necessarily real) are examined.  The determining of
these principles should precede to systematic examination of wave equations
in concrete spaces. The plan of the present work is the following.  In \S 2
the connection of wave equations with the symmetry group of space is
explained without concrete definition of it. It is shown
how it is possible to obtain the wave equations in symmetrical curved space
via the decomposition of its symmetry group representation which describes
particles with  spin in the direct product of scalar and purely matrix
representations.  However, both curved space and the invariant wave equations
on it should obey  the {\it principle of the correspondence}, i.e. admit a
 passage in the flat space limit to the usual relativistic
wave equations following
from its symmetry group.  In \S 3 it is shown how with the help of known
physically important wave operators in the flat (not only real) spaces it is
possible to obtain:
\begin{enumerate}
\item[]{(a)}
curved symmetrical spaces in which the
invariant wave equations  obeys the principle of the correspondence,
\item[]{(b)} their symmetry groups.
\item[]{(c)} their matrix representations which are needed for the invariant
wave equations on this spaces construction.
\end{enumerate}
In \S 4 the method of \S 3 is carried out for real spaces starting from the
Dirac operator  and from the wave equations for
massless particles of an arbitrary spin in  the Minkowski space. The list of
real spaces which admits the group theoretical examination of wave
equations is obtained. In \S 5 this procedure is carried out for complex
spaces, the corresponding wave equations  already finding the numerous
 physical applications~\cite {26}; concerning the Dirac operator on
complex manifolds see also~\cite{50} and references therein.
It is also necessary for this to consider
 the wave equations on the flat space ${\Bbb C}^{n} $. It is shown that the
usual Dirac equation written down in the complex variables does not lead  to
any curved space. By "decomposition" of the mentioned form of the Dirac
equation and doubling  its matrix dimensionality the equation leading to the
space ${\Bbb C}{\Bbb P}^{n}$ and being alternative to the usual  Dirac
equation on space ${\Bbb C}^{n}$ is constructed.

The following paper will be devoted the concrete examination of wave equations
in the found real irreducible spaces.

\section {The symmetry of space and wave equations }
Let space have some continuous symmetry group~${\cal G} $.  
As in the case of the Minkowski space~\cite{18,1}, it is natural to
connect elementary particles with the irreducible representations of the
group~${\cal G}$.
After the irreducible representation of the group ${\cal G}$ is fixed, the
wave equations for particles, described by this representation, can
be obtained unambiguously as follows~\cite {3}. Let us consider an operation
of the group ${\cal G} $ on the configuration space; it will give us the
representation of its generators as some differential operators $-$ as the
generators of the {\it scalar representation} of the group ${\cal G}$. Then
we shall add to this  {\it orbital} part a certain constant matrix ({\it
spin} part) which corresponds to the generators of some finite-dimensional
matrix representation of the group~${\cal G}$. It should be chosen so that
the summarized generators compose the representation  describing our
particle.  In the language of group representations  it means decomposition
of the irreducible representation of symmetry group into the direct product
of scalar and purely matrix representations.

Such a decomposition of generators has the simple physical sense. Really, the
change of the wave function under the group transformations is
$$\psi'(x)-\psi(x)=-(\psi'(x') -\psi'(x))+(\psi'(x') -\psi(x)). $$
The first bracket in  right hand side corresponds to the change of the
observation point;  the orbital part of generators is responsible for it.
The second bracket corresponds to the transformations of the wave function in
the same  point; it is obvious that such transformations will form a group.
For them responsible are the spin parts of generators, which, thus, will form
a representation of the group~${\cal G}$.  Now form the operators which
commute with all generators of the representation (if these operators are
constructed only from the group generators, then they are called the {\it
Casimir operators}) .  The first Schur's lemma states that such operators
have the fixed value in the irreducible representation. They will give us the
wave equations.

How does the first-order wave equations for particles with the nonzero spin
follow from the group ${\cal G}$?  The Casimir operator  $C_{2} ({\cal G})$
 by definition is quadratic in generators. Therefore, by substituting in it
the generators as the sum of orbital and spin parts we obtain
$$C_{2}=C^{(l)}_{2}+C^{(l-s)}_{2}+C^{(s)}_{2}. $$
The first and third terms in the right hand side  are Casimir operators of
scalar and matrix representations. It is clear that they commute with the
orbital and spin parts of generators separately, and, so, with full
generators of representation of the appropriate spin.  Therefore, the same is
correct also for the second term  $C^{(l-s)}_{2}$,  which
is on the first power of both the orbital
and  spin parts of generators.  It is just  (up to
a constant) the wave operator, and its eigenvalues will give eigenvalues of
this operator.

\section {Method of admissible symmetric spaces obtaining}

For the meaning of the group theoretical examination of the elementary
particle in the curved space (let us  designate it $\cal M$) it is
necessary first, that it allow the limiting passage on flat space.
I.e.  its metric should include the parameter $R$ which has the sense
 of the curvature radius. When $R$ tends to infinity, the metric should
pass to the Galilean metric on the flat space (we shall designate it
 ${\cal M}'$). Secondly, the principle of the correspondence for the
invariant wave equations on $\cal M$ should be satisfied i.e. they
 should pass in the limit $R\rightarrow\infty$ into the some reasonnable
wave equations on ${\cal M}'$ following from some representation of its
symmetry group~${\cal G}'$.

To obtain the spaces and invariant wave equations on them which
obey  these requirement we shall act as follows.

Let the symmetry group ${\cal G}'$ of the flat space is known.  Let us take
now  the symmetry group $\cal G$ (not yet known) of curved space and tend
$R$ to infinity.  Then the Casimir operator $C_{2}({\cal G})$ will  coincide
 within $o(\frac{1}{R})$ terms with the known operator $C_{2}({\cal G}')$.
 Further, the orbital parts of generators will coincide with the same
 accuracy with those for the group ${\cal G}'$, which also are known.
Therefore, the orbital-spin part $C_{2}^{(l-s)}({\cal G})$ can be expressed
through differential operators on the flat space and while unknown spin parts
of generators (probably, of not all)  of group $\cal G$. The comparison of
this expression with  known wave equations for the flat space  allow us to
obtain the mentioned spin parts.  Making the commutators between them
allows one to obtain the rest of the  generators
of the group $\cal G$ matrix representation and the commutation relations of
this group.  In general, this procedure can lead to the not quite unambiguous
results, and for more precise definition it is necessary to make sure that
the obtained curved space really supposes the limiting passage  to the flat
space. For it the correctness of the condition
 $$\dim {\cal M} = \dim {\cal M} ' $$
is necessary. Let us remark that any symmetrical space which is not  a Lie
 group (we shall consider only such spaces)  is possible to present as the
coset space~\cite {29}:
$${\cal M} = {\cal G}/{\cal H} \quad , \quad
{\cal M}' = {\cal G}'/{\cal H}',$$
where ${\cal H} \subset {\cal G}$ ($ {\cal H} '\subset {\cal G}'$) is the
stationary subgroup of an arbitrary point of $\cal M $ (${\cal M} ' $).
Since in the infinitesimal neighborhoods of $\cal M$ and ${\cal M}'$
should be isomorphic (in the opposite case the limiting passage from $\cal M$
to  ${\cal M}'$ would not be possible) and $\cal H$ is the maximal stationary
subgroup for such spaces then
$$\dim {\cal H}\leq \dim {\cal H}'.$$
On the other hand, the real
dimensionality of $\cal M $ is equal to
\begin{equation}\label{2. a2}
\dim {\cal M} = \dim {\cal G} -\dim {\cal H}
\end{equation}
and  analogously for ${\cal M} ' $.
Therefore,
\begin{equation} \label {2. a1}
\dim {\cal G} \leq \dim  {\cal G} '.
\end{equation}

\section {Admissible real spaces}
Let us consider how the above procedure may be performed for the  real spaces.
Taking into account the physical importance of four-dimensional
space of the Lorentz signature, we shall consider just this case.

Within higher orders on $1/R$,
the Casimir operator  $C_{2}$ of the
group ${\cal G}$ coincides with that of the Poincar\'e group:
${\cal G}'={\cal P} \equiv {\Bbb R}^4 \x SO (3, 1)$
(where $\x$ is the semidirect product):
$$C_{2}({\cal G}) \approx C_{2}({\cal P})=P^{\mu}P_{\mu}, $$
where $P_{\mu}$ are the generators of translations. Therefore
$$C^{(l -s)}_{2}({\cal G}) \approx 2P^{(s)\mu}P^{(l)}_{\mu}.  $$
Since
$$P^{(l)}_{\mu} \approx \partial_{\mu}, $$
  the first-order wave operator  in the Minkowski space is
\begin{equation}\label{2. 1}
\mbox{const}\cdot P^{(s) \mu} \partial_{\mu}.
\end{equation}
Let us compare this to
the equations for massive and massless particles following from
the Poincar\'e group by considering these cases separately.

1) The equations for massless particles of an arbitrary spin have the
form~\cite{3}
 \begin{equation}\label{2. 2}
(W_{\mu}+i\lambda P_{\mu \ {\rm Poinc}}) \psi=0,
\end{equation}
where $W_{\mu}$ is the Pauli-Lubanski pseudovector:
$$W_{\mu}=\frac{1}{2}
\varepsilon_{\nu \rho \sigma \mu} J^{\nu \rho}P^{\sigma}_{{\rm Poinc}}, $$
$\lambda$ is helicity: $\lambda = \pm s,\ s=1/2,1,\ldots$
Operators $P_{\mu \ {{\rm Poinc}}}$ and
$J_{\mu \nu}$ are the generators of the appropriate representation of the
Poincar\'e group:
\begin{equation}\label{2.  4}
\begin{eqalign}
P_{\mu \ {{\rm Poinc}}}=\partial_{\mu} \\
J_{\mu \nu}=x^{\rho}(\eta_{\mu \rho}\partial_{\nu}-
\eta_{\nu \rho}\partial_{\mu})+J^{(s)}_{\mu \nu},
\end{eqalign}
\end{equation}
where $J^{(s) \mu \nu}$ are the generators of  the Lorentz group
representation $L^{(s,0)}$
(if $\lambda =s$) or $L^{(0, s)}$ (if $\lambda =-s$), $\eta_{\mu
\nu}=\mbox{diag}(+ - - -)$ is the Galilean metric tensor. Since for them
$$\frac{1}{2} \varepsilon_{\mu \nu \rho \sigma}
J^{(s) \rho \sigma}=\mp J^{(s)}_{\mu \nu}, $$
then~(\ref{2. 2})  may be simplified:
\begin{equation}\label{2. 5}
(s\eta^{\mu \nu}+J^{(s)\mu \nu}) \partial_{\nu} \psi=0.
\end{equation}
The set of equations~(\ref{2.  5}) is surplus, hence its temporal component
is sufficient to obtaining  the form of generators
$P^{(s) \mu}$.
As
$$J^{(s)i0}=\mp iX^{(s)i}, $$
where $X^{(s)i} =\frac{1}{2} \varepsilon^{ikl} J^{(s)kl}$ are the
generators of the appropriate
representation of  rotation group, we finally obtain:
\begin{equation} \label{2. 6}
( s\partial_{t} \pm iX^{(s) i}\partial_{i}) \psi=0.
\end{equation}
Comparing this with~(\ref{2. 1}) it follows
\begin{equation}\label{2.  7}
\begin{eqalign}
 P^{(s)0}=i \lambda /R \\
 P^{(s)i}=X^{(s)i}/R.
 \end{eqalign}
 \end{equation}
The commutation relations between
these generators will give us the group $T_{1}^{\Bbb R} \otimes
SO(3)$; the space which correspond to it donesn't have a conventional name.
This group may be  supplemented up to the group ${\Bbb R}^1 \otimes SO(4)$ 
by the spatial rotations; it will give the Einstein space.

If $R$ in~(\ref{2. 7}) replaced by $iR$, the generators will not
compose any real group; being supplemented with rotations they will compose
the group ${\Bbb R}^1 \otimes SO (3, 1)$. The corresponding space is the
Anti-Einstein space ${\Bbb R}^1 \otimes H^3$, where $H^3$ is the
three-dimensional hyperbolic space.

2) Massive representation of spin 1/2. The comparison of the Dirac equation
 $$ (i\gamma^{\mu} \partial_{\mu}-m) \psi=0 $$
with~(\ref{2. 1})  yields
$$ P^{(s) \mu}=\mbox{const} \cdot \gamma^{\mu}. $$
Taking this together with their commutators, which are the
generators of   the Lorentz transformations, we shall come to the groups
$SO(4,1)$ or $SO(3,2)$. Spaces supposing such symmetry groups  are the de
Sitter and Anti-de Sitter spaces.

So,  the following real spaces are subject to examination:
\begin{enumerate}
\item Reducible spaces
\begin{enumerate}
\item Einstein space ${\Bbb R}^{1}\times S^{3}$ ,
${\cal G}={\Bbb R}^1 \otimes SO (4) $ \\
\item Anti-Einstein space ${\Bbb R}^1 \otimes H^3$, \
${\cal G}={\Bbb R}^1 \otimes SO (3, 1) $. \\
\item ${\cal G}= {\Bbb R}^1 \otimes SO(3),
 \quad {\cal H} $ is trivial.
\end{enumerate}
\item De Sitter and Anti-de Sitter spaces.
\begin{enumerate}
\item De Sitter space  $SO(4,1)/SO(3,1)$. \\
\item Anti-de Sitter space   $SO(3, 2)/SO(3,1)$.
\end{enumerate}
\end{enumerate}

\section {Admissible complex spaces}
Let us designate the coordinates of the $n$-dimensional
 complex space through $z_{i},\ i=1,\ldots,n $; the complex conjugation is
designated by the over-line.
The symmetry group of the flat complex space with the metric
$$ds^{2}=dz_{i}d{\overline z}_{i}$$
be $T_{n}^{\Bbb C} \x U(n)$
with the Hermitian generators $K_{ik}$ and $J_{ik}$
which may be constructed in the following form:
\begin{equation}\label{A7a}
\begin{eqalign}
J_{ik}=A_{ik}-A_{ki}, \\
K_{ik}=i(A_{ik}+A_{ki}).
\end{eqalign}
\end{equation}
Where $A_{ik}$  are the generators of the group
$GL(n,{\Bbb C})$ which have the property
\begin{equation}\label{A9}
(A_{ik})^{\dagger}=A_{ki}
\end{equation}
and obey the commutation relations
\begin{equation}\label{A5}
[A_{ik}, A_{lm}]=\delta_{kl}A_{im}-\delta_{im}
A_{lk}.
\end{equation}
 As, according to~(\ref{A7a}), the Jacobian of the
passage from the group $GL(n,{\Bbb C})$ generators to the group $U(n)$
generators is not equal to zero, we can, that  is much more convenient,  deal
with the generators $A_{ik}$.

We have:
$$C_{2}({\cal G}')=C_{2}(T_{n}^{\Bbb C}) =
-P_{i} P^{\dagger}_{i}.$$
The scalar representation  of the translation group be
$$P^{(l)}_{i}=\partial_{i} \ ,  \ P^{(l)\dagger}_{i}=
-{\overline \partial}_{i}, $$
where $\partial_{i}=\frac{\partial}{\partial z_{i}}$ . Note that
$(\partial_{i})^{\dagger}=-{\overline \partial}_{i}$. Therefore
\begin{equation}\label{2. 8}
 C_{2}^{(l-s)}({\cal G}) \approx P^{(s)}_{i}{\overline \partial}_{i} -
P^{(s)\dagger}_{i}\partial_{i}.
\end{equation}
Let us obtain the wave equations following from the
group ${\cal G}'$. The generators of scalar and "spinor"  representations of
the group $GL(n,{\Bbb C})$  are
$$A^{(l)}_{ik}=z_{i} \partial_{k} -
{\overline z}_{k}{\overline \partial}_{i}$$
\begin{equation} \label{2. 9}
A^{(s)}_{ik}=b_{i}b_{k}^{\dagger}+ a\delta_{ik},
\end{equation}
Where $a$ is an arbitrary real constant.  It is easy to show
that they obey commutation relations~(\ref{A5}).
  Here $b_{i}$ are the fermionic operators of birth and annihilation:
\begin{equation}\label{2.  10}
\{ b_{i},b_{k} \} = \{ b^{\dagger}_{i},b^{\dagger}_{k} \} =0 \quad,\quad
\{ b_{i},b^{\dagger}_{k} \} =\delta_{ik},
\end{equation}
which are realized with the help of $2^{n}$-dimensional
matrices. By  using them we can construct
 the $\gamma$-matrices in $2n$ dimensions:
\begin{equation}\label{2.  11}
\gamma_{k}=b_{k}+b_{k}^{\dagger} \quad,\quad
\gamma_{n+k}=-i(b_{k}-b^{\dagger}_{k}).
\end{equation}
So $\{ \gamma_{p}, \gamma_{q} \} =2\delta_{pq}$, $p,q=1,\ldots,2n.$
 It is possible to construct the matrix
$$\gamma_{2n + 1} =i^{n(2n-1)} \gamma_{1} \gamma_{2} \ldots \gamma_{2n}, $$
commuting with all generators $A^{(s)}_{ik}$ and having an eigenvalue
$\pm 1 $. Thus, the representation~(\ref{2. 9}) is two-multiple reducible.
Note that if the additional  condition $A^{(s)}_{ii} \psi=0$ is imposed, the
representation of the group $U(n)$ with weights $1/2(\pm 1, \pm 1,
\ldots, \pm 1)$ may be obtain, where the signs in each case are chosen
independently.  Now, construct the operator
$$H=b_{i} \partial_{i}. $$
Using~(\ref{2. 10}) it is easy to show that the operators $H$
and $H^{\dagger}$ commute with all summarized generators $A_{ik}$ and
consequently are acceptable for the construction of wave equations. Note that
$$H-H^{\dagger} = \gamma_{p} \frac{\partial}{\partial x_{p}}, $$
where $z_{k} = x_{k}+ix_{k + n} $, is the Dirac
operator on ${\Bbb R}^{2n}$.  This corresponds to an embedding $U(n)$ in
$SO(2n)$. The formulas  following from~(\ref{2. 10})
$$\begin{eqalign}
H^{2}=(H^{\dagger})^{2}=0 \\
(H-H^{\dagger})^{2}=\Box \equiv \partial_{i}{\overline \partial}_{i}
\end{eqalign}$$
yields that
the Dirac operator  is the only operator of matrix dimensionality
$2^{n}$, which would be constructed from operators $H$ and $H^{\dagger}$ and
 leads by  squaring to the Klein-Gordon operator, if we want to
describe massive particles (so, the {equation} $H\psi =H^{\dagger} \psi=0$
 is excluded).

The Dirac operator written down in the complex variables is the conventional
spinor wave operator on complex manifolds~\cite{26}.
  However, if we start from this operator using the method of \S 4,
then we obtain $2n^{2}+n$ generators
$b_{i}$, $b_{i}^{\dagger}$,
$[b_{i},b_{k}^{\dagger}]$, $b_{i}b_{k}$ and $b_{i}^{\dagger}b_{k}^{\dagger}$
in the contradiction with~(\ref{2. a1}). Thus, the obtained space does not
suppose the limiting passage to ${\Bbb C}^{n}$.  Besides, this operator,
having "surplus" symmetry for our case, is not specific for the flat complex
 space.

One can construct
the wave operator does not have these shortages, as
follows. Let us construct four-multiple reducible representation
\begin{equation} \label{2. 12}
A^{(s)}_{ik} =(b_{i} b_{k}^{\dagger} +a\delta_{ik}) \cdot\left(
\begin{array}{ll}
 1 & 0 \\
 0 & 1
\end{array} \right).
\end{equation}
Then the operator
\begin{displaymath} D =
\left( \begin{array}{ll}
 0 & H^{\dagger} \\
 H & 0
\end{array} \right),
\end{displaymath}
as well as the operator ${\tilde D}=SDS^{-1}$, where
$$S=S^{-1}=\left( \begin{array}{cc}
0 & 1 \\
1 & 0
\end{array} \right), $$
will commute with new summarized generators
$A^{(l)}_{ik} + A^{(s)}_{ik}$. Since $C_{2}^{(l)} ({\cal G}')=\Box$, we
have the second-order equation
$$(\Box +\alpha^{2}) \psi=0. $$
The transition from the first-order wave operators  to  the second-order is
carried out either with the  equalities
\begin{equation}\label{2.  12a}
\begin{eqalign}
D{\tilde D}={\tilde D}D=0 \\
D^{2}+{\tilde D}^{2}=-\Box
\end{eqalign}
\end{equation}
or with the equality
$$D^{3} =\Box D. $$

There are two kinds of first-order wave equations  compatible
with~(\ref{2.  12a}). From each of them it follows
an additional wave equation:
\begin{equation} \label{2. 13}
D\psi =\alpha\psi \quad (\Rightarrow{\tilde D} \psi = 0),
\end{equation}
\begin{equation} \label{2. 14}
{\tilde D} \psi=\alpha\psi \quad (\Rightarrow D\psi = 0).
\end{equation}
This, together with ambiguity in a sign $\alpha$, reduces to the
four-multiple ambiguity; the choice  of  one of these possibilities
compensates\footnote{It is similar to the obtaining of the Dirac equation
from the Poincar\'e group.  In this case the two-multiple ambiguity in an
eigenvalue of the Dirac operator compensates the two-multiple  reducibility
of the Lorentz group matrix representation~\cite{3}.} the four-multiple
reducibility of the representation~(\ref{2. 12}) and leads us to the
irreducible representation of group ${\cal G}'$.  We will  choose the
equation~(\ref{2. 13}). Comparing the
equation~(\ref{2. 13}) with~(\ref{2. 8}) gives
\begin{displaymath}
 R\cdot P^{(s)}_{i}=\left(
\begin{array}{ll}
 0 & b_{i}^{\dagger} \\
0 & 0
\end{array} \right).
\end{displaymath}
 The obtained generators, together with~(\ref{2. 12}), can be
closed in the Lie algebra by putting
\begin{displaymath}
\begin{eqalign}
A_{0i}=R\cdot P_{i}, \\
A_{00}=\left(
\begin{array}{ll}
  a+1 & 0 \\
0 & a
\end{array}
  \right) .
\end{eqalign}
\end{displaymath}
They obey the commutation relations of the group $GL(n+1, {\Bbb C})$:
\begin{equation} \label{2. 15}
[A_{\mu\nu}, A_{\rho\sigma}] = \delta_{\nu\rho} A_{\mu\sigma} -
\delta_{\mu\sigma} A_{\rho\nu},
\end{equation}
where $\mu,\nu,\ldots=0,\ldots,n$.

Starting from the equation~(\ref{2. 14}) yields the generators
$$\begin{eqalign}
R\cdot {\tilde P}^{(s)}_{i}=\left(
\begin{array}{ll}
0 & 0 \\
 b_{i}^{\dagger} & 0
\end{array}
  \right) \\
 {\tilde A}_{00}=\pm\left(
\begin{array}{ll}
  a & 0 \\
0 & a+1
\end{array}
  \right)  \\
{\tilde A}_{ik}=A_{ik},
\end{eqalign}$$
which also obey  the commutation relations~(\ref{2. 15}). Note
that two spinor representations, which we have constructed, are
equivalent to each other, since
$$ {\tilde A}_{\mu\nu} = SA_{\mu\nu} S^{-1}. $$
In general,  all these considerations require a verification since the
 symmetry group of flat space is $U (n)$ but not $GL (n,{ \Bbb C})$.
However, generators of  $U (n + 1)$ are  constructed from
generators of  $GL (n + 1, {\Bbb C})$ in  the same way as
the generators of $U (n)$ are constructed from the generators of $GL
(n, {\Bbb C})$. Therefore, if we started from $U (n)$ then we would come to
the group $U (n + 1)$ and the symmetrical space $U (n + 1) /U (n)$. It,
however, does not suppose the limiting passage to ${\Bbb C}^{n}$ because
its real dimensionality calculated with the help of~(\ref{2. a2}) equals
$2n + 1$ in the contradiction with~(\ref {2. a1}). Therefore, we demand
that the operator $A_{\mu\mu}$ with the operation on an arbitrary
vector of representation space give  zero. Thus, we shall pass to group $SU
(n + 1)$ and complex projective space
$$ {\Bbb C}{\Bbb P}^{n} = SU (n + 1) /U (n)$$
which already supposes the limiting passage  to ${\Bbb C}^{n}$.

I am grateful  to Yu.P.Stepanovsky for the constant and versatile support
during the work.

\end{document}